\documentclass{nature}

\usepackage{graphicx,subfigure,epsfig,epstopdf}
\usepackage{amsmath,amssymb}
\input{epsf}
\input{epsfx}

\def\bra#1{\mathinner{\langle{#1}|}}
\def\ket#1{\mathinner{|{#1}\rangle}}
\def\braket#1{\mathinner{\langle{#1}\rangle}}

\newcommand{\unit}[1]{\hspace{-1.3pt} {#1}}

\begin{document}
\title{Coherent excitation of a strongly coupled quantum dot - cavity system}

\author{Dirk Englund$^{1,2}$,  Arka Majumdar$^{1}$, Andrei Faraon$^{1,3}$, Mitsuru Toishi$^{4}$, Nick Stoltz$^{5}$,   Pierre Petroff$^{5}$ \& Jelena Vu\v{c}kovi\'{c}$^{1}$}

\maketitle
\begin{affiliations}
 \item[1. ] Department of Electrical Engineering, Stanford University, Stanford CA 94305
 \item[2. ] Department of Physics, Lyman Laboratory, Harvard University, Harvard MA 02138
 \item[3. ] Department or Applied Physics, Stanford University, Stanford CA 94305
 \item[4. ] Sony Corporation, Shinagawa-ku, Tokyo, Japan, 141-0001
 \item[5. ] Department of Electrical and Computer Engineering, University of California, Santa Barbara, CA 93106
\end{affiliations}

\maketitle

\begin{abstract}
Photonic nanocavities coupled to semiconductor quantum dots are becoming well developed systems for studying cavity quantum electrodynamics and constructing the basic architecture for quantum information science.  One of the key challenges is to coherently control the state of the quantum dot/cavity system for quantum memory and gates that exploit the nonlinearity of such a system\cite{1999.PRL.Imamoglu-Small.QIP-QD,1999.PRL.Haroche.phase-gate,2003.Science.Li-Scham.QGate,Sham2005PRL}.  Recently, coherent control of quantum dots has been studied in bulk semiconductor \cite{2007.Science.Xu-Steel,2008.PRL.Jundt-Imamoglu.Dressed_QD_states_resonant,2008.Nature.Press.complete_quantum_control}.  Here we investigate the coherent excitation of a strongly coupled InAs quantum dot - photonic crystal cavity system. When the quantum dot and cavity are on resonance, we observe time-domain Rabi oscillation in the transmission of a laser pulse. This coherent excitation promises to enable an all-optical method to observe and manipulate the state a single quantum dot in a cavity.   When the detuned dot is resonantly excited, we show that the resonantly driven quantum dot efficiently emits through the cavity mode, an effect that is explained in part by an incoherent dephasing mechanism similar to recent theoretical models\cite{2008.PRL.Mork.dephasing,2008.OpEx.Noda.dephasing,2008.ArXiv.Poizat.dephasing}.  When the detuned quantum dot is resonantly excited, the cavity signal represents a spectrally separated read-out channel for high resolution single quantum dot spectroscopy. In this case, we observe antibunching of the cavity mode. Such a single photon source could allow photon indistinguishability that approaches unity\cite{2008.ArXiv.Poizat.dephasing} as could lift the limitation due to dephasing and timing jitter\cite{2002.Nature.Santori.indist-phot,2004.PRA.Kiraz.SPS-prospects}. The single photon emission is controlled by the cavity resonance, which relaxes the demands for spectrally matching quantum dots for two-photon interference and may therefore be of use in linear optics quantum computation\cite{KLM01} and quantum communication \cite{2003.PRL.Simon-Irvine.entanglement_indist_photons,2005.PRA.Childress-Lukin.quantum-repeater-SP}. 
\end{abstract}


\begin{figure}
\centering{\includegraphics[width=.8\linewidth]{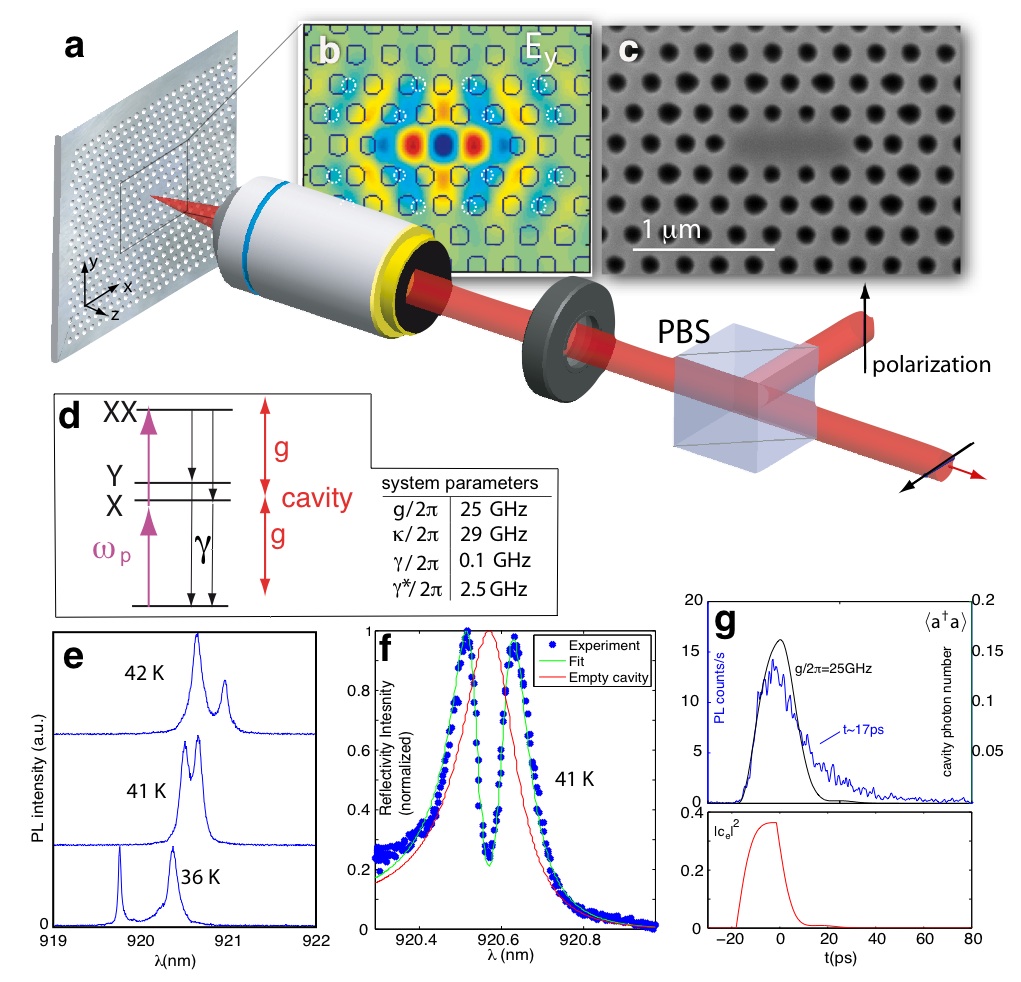}}
\caption{{\footnotesize (a) Cross-polarized confocal microscope setup.  (b) $E_y$ field in the L3-cavity structure. The structure is perturbed at the sites indicated by dashed circles, as described in Ref.\cite{2008.OpEx.Englund}. (c) Scanning electron micrograph of fabricated structure. (d) Energy levels of the coupled QD/cavity system showing the two polarization states of the single exciton (X,Y) and the bi-exciton state (XX). The table lists the system parameters derived from the measurements, where $g,\kappa,\gamma,\gamma^{*}$ are the vacuum Rabi frequency, cavity field decay rate, dipole decay rate, and dipole dephasing rate, respectively. (e) The photoluminescence (PL) shows the QD/cavity anticrossing as the QD is temperature-tuned through the cavity.  (f) Vacuum Rabi splitting observed in the reflectivity from the strongly coupled QD/cavity system. For comparison, we show the reflectivity of an empty cavity. (g) PL lifetime $\sim 17$\unit{ps} when the QD is tuned into the cavity and excitation wavelength $\lambda_p=878$\unit{nm}. The emission that is expected theoretically, based on the system parameters in (d) and a 10-ps relaxation time into the single exciton state, is shown in the solid line. The bottom panel plots the corresponding expected excited state population $|c_e(t)|^{2}$.} }\label{fig:setup}
\end{figure}

The optical system consists of a photonic crystal (PC) cavity fabricated in a 160-nm thick GaAs membrane by a combination of electron beam lithography and dry/wet etching steps, as discussed in Ref.\citen{2005.PRL.Englund}.  The membrane contains a central layer of self-assembled InGaAs quantum dots (QDs) with an estimated density of $50/\mu$m$^{2}$.  The completed photonic crystal is shown in the scanning electron micrograph in Fig.\ref{fig:setup}(c). 

We employ a grating-integrated cavity (GIC) design for improved input and output coupling\cite{2008.OpEx.Englund}. The cavity design is based on a linear three-hole defect cavity\cite{Noda2003Nature}. The perturbed cavity design increases the directionality of the unperturbed cavity emission and improves the in- and out-coupling efficiency\cite{2008.OpEx.Englund}. The integrated grating has a second-order periodicity as indicated in Fig.\ref{fig:setup}(b). 

We first characterize the QD/cavity system by its photoluminescence.  The sample is maintained between 10\unit{K} and 50\unit{K} in a liquid-He continuous flow cryostat. A continuous-wave (cw) laser beam at 860\unit{nm} excites electron-hole pairs  which can relax through a phonon-mediated process into radiative levels of the QD. This above-band driving case corresponds to the pump frequency $\omega_p$ tuned above the single-exciton frequency that is labeled as $X$ in Fig.\ref{fig:setup}(d). Fig.\ref{fig:setup}(e) plots the photoluminescence spectrum.  As we sweep the temperature, we observe the anticrossing between the QD emission and the cavity emission that is characteristic of the strong cavity-emitter coupling.  From the QD/cavity spectrum, we estimate the system parameters summarized in the table in Fig.\ref{fig:setup}(d). 

Next, we resonantly excite the QD by coupling a tunable cw-laser beam into the cavity. The laser has a linewidth below $ 300$\unit{kHz}. Using the cross-polarized setup illustrated in Fig.\ref{fig:setup}(a), we observe the transmission of the incident vertically polarized laser via the cavity into the horizontally polarized component.  We clearly observe the vacuum Rabi splitting\cite{2007.Nature1,1992.PRL.Thompson-Kimble.Normal_mode_splitting,2006.PRL.Waks-Vuckovic.DIT}, which demonstrates that we are coherently probing the QD/cavity system. The reflectivity signal nearly vanishes when the laser field is resonant with the QD single exciton (X) frequency, showing that the QD has a very high probability of being in the optically bright state. We obtain good agreement with theory (solid line fit in Fig.\ref{fig:setup}(f)). In this fit, we used the same parameters as derived from the photoluminescence (PL) data in Fig.\ref{fig:setup}(e). When the quantum dot is on resonance with the cavity and is pumped at 878\unit{nm}, the PL decays with a characteristic time of 17\unit{ps}. The decay time matches a theoretical model based on the system parameters, shown in the solid line in Fig.\ref{fig:setup}(g).  The model considers a quantum dot which is pumped into the single-exciton excited state (including timing jitter) and then decays into free space and the cavity mode (see Methods).  The state of the QD is described by a superposition of ground and excited states, $\ket{\psi(t)}=c_g(t) \ket{g} + c_e(t) \ket{e}$. The expected excited state population $|c_e(t)|^{{2}}$ that is predicted by the fitting model is plotted in the bottom panel of Fig.\ref{fig:setup}(g). Previous measurements of the decay time gave values exceeding 60\unit{ps} in the strong coupling regime\cite{2007.Nature.Hennessy-Imamoglu.Strong_coupling_quantum_nature}, which is longer than expected for the strong coupling regime where the decay time should be on the order of the cavity ring-down time of $\sim 5$\unit{ps} for a $Q\sim 10^4$. We attribute the short decay time in this case to the observation that nearly all emission collected from the cavity originates from the QD.  For a very similar system, we previously showed that the cavity mode is strongly antibunched to ($g^{(2)}(0)\sim 0.05 $) \cite{2008.APL.Toishi_Englund}. The perturbed cavity lifts the QD signal far above the background and thus eliminates the collection of the long lived emission lines that would degrade antibunching.

\begin{figure}
\centering{\includegraphics[width=\linewidth]{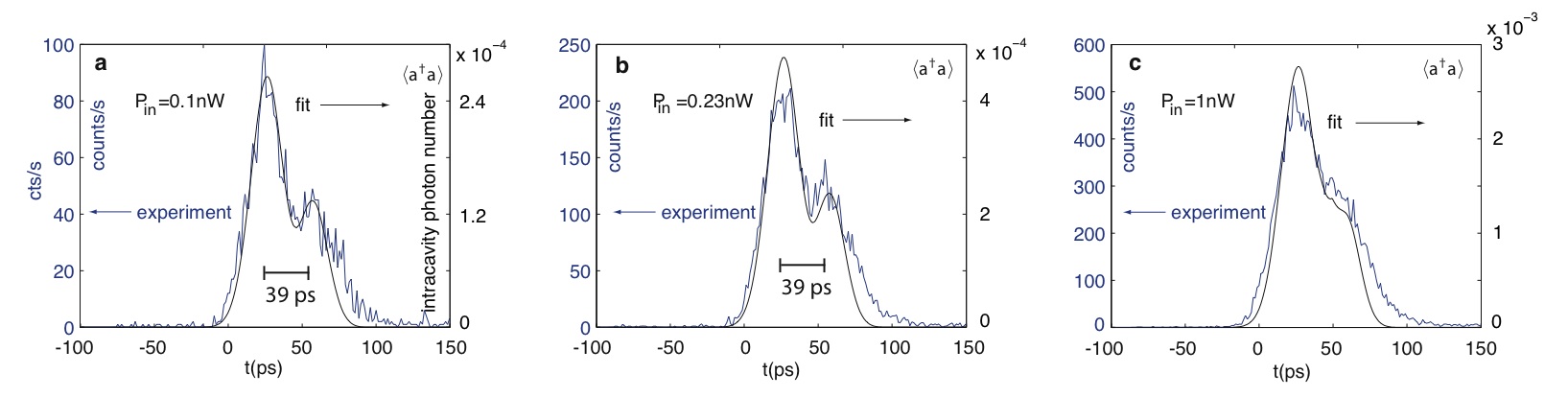}}
\caption{{\footnotesize Time-resolved reflectivity measurement shows Rabi oscillation frequency $g/2\pi = 29 $GHz.  We show three excitation powers in (a,b,c).  The observations are fit with a full master equation model using the parameters given above; in all three plots, the photon flux into the cavity is scaled equally from the measured incident power (0.1nW, 0.23 nW, 1 nW before the lens). }} \label{fig:time_domain} 
\end{figure}

In the PL and reflectivity measurements described above, we characterized the QD/cavity in the frequency regime. We will now describe a method to observe the system dynamics by direct time domain measurements of the vacuum Rabi frequency. Instead of the cw probe laser, we now use a spectrally filtered Ti:Sapphire laser with 40\unit{ps} pulses at 80 MHz repetition rate. Since the reflected beam intensity is weak when detuned from the cavity (as evident from Fig.\ref{fig:setup}(f)), we performed this measurement when the QD was resonant with the cavity, although a range of detunings and vacuum Rabi frequencies are in general possible. In Fig.\ref{fig:time_domain}, we plot the time-resolved reflected pump intensity, which is acquired with a streak camera (see Methods). The period of $39$\unit{ps} closely matches the expected Rabi period $T=2\pi/g = 40$\unit{ps}.  At higher pump power, the Rabi oscillation becomes less visible (Fig.\ref{fig:time_domain}(c)).  In all observed cases, we found good agreement with theory (plotted in solid lines), which is obtained using a quantum Monte Carlo simulation (see Methods). The fits assume the values of $g,\gamma,$ and $\kappa$ obtained from spectral measurements, i.e., PL and reflectivity.  The time-resolved reflectivity presented here offers a direct tool for coherently manipulating the state of the QD/cavity system. 

\begin{figure}
\centering{\includegraphics[width=1\linewidth]{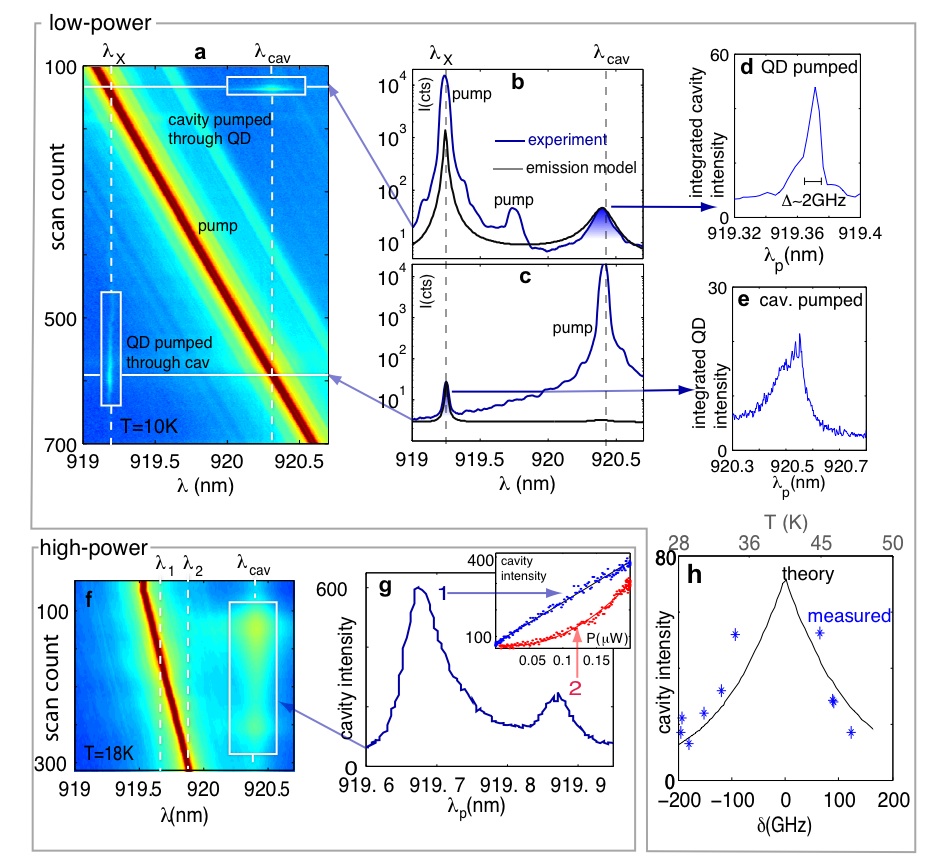}}
\caption{\footnotesize{Coherent QD spectroscopy through cavity mode. (a) Intensity on spectrometer when the pump wavelength $\lambda_p$ is scanned across the detuned QD/cavity system (the main laser excitation is accompanied by a red-detuned side mode that is 200 times lower in intensity; temperature is 10K).  When $\lambda_p$ is resonant with the dot (cavity), the cavity (QD) intensity rises. (b) Spectrum when QD is pumped (the middle peak corresponds to the laser side mode) (c) spectrum when cavity is pumped. In (b) and (c), we model the driving mechanism by a pure dephasing process with $\gamma^{*}=0.1 g$. The model agrees well with the emission through the cavity (QD) when the QD (cavity) is pumped.  (d) The integrated cavity emission as a function of the pump wavelength $\lambda_p$ shows the single exciton absorption resolved to $3$\unit{GHz}. (e) The integrated QD emission shows a non-lorentzian dependence on the pump frequency into the cavity. (f) At high pump power, a second line becomes visible at $\lambda_p=919.87$\unit{nm}. (The lines are slightly shifted since the temperature was, for technical reasons, raised to 18 K). (e) Integrated cavity emission as $\lambda_p$ is scanned. \textit{Inset:} The power dependence of the two lines suggests exciton (X) and bi-exciton (XX) states. (g)  Dependence of integrated cavity emission on QD detuning when the QD is resonantly pumped.}} \label{fig:coherent_CW}
\end{figure}

We will now consider the resonant driving of the quantum dot when it is detuned from the cavity and will demonstrate the use of the cavity mode as a convenient read-out and pumping channel for resonant single quantum dot spectroscopy and single photon generation. We first lower the temperature to 10K, which blue-detunes the dot by $\delta=\lambda_{d}-\lambda_{c}=-1.17$\unit{nm} from the cavity resonance. The laser excitation is polarized at 45$^{\circ}$ to the cavity mode, where its alignment can be optimized for the reflectivity signal shown in Fig.\ref{fig:setup}(f). Then we scan the laser across the QD and cavity resonances, as shown in Fig.\ref{fig:coherent_CW}. The excitation laser power is $\sim 12$\unit{nW} before the objective lens. Precisely when the laser becomes resonant with the QD, we observe a strong emission into the cavity mode. Thus the cavity represents a strong read-out channel for resonant quantum dot spectroscopy: the resonantly driven QD emits into the cavity mode, which is far detuned and easy to separate spectrally. Alternatively, if the cavity mode is pumped, the QD single-excition line radiates. Figs.\ref{fig:coherent_CW}(b,c) plot spectra when the (QD,cavity) are pumped. 

Fig.\ref{fig:coherent_CW}(d) plots the integrated cavity emission as a function of the laser pump wavelength $\lambda_p$.  The QD absorption linewidth is measured to be lower than $0.006$\unit{nm} (2 GHz) -- about five times narrower than the $\sim 0.03$\unit{nm} resolution of our $0.75$ m spectrometer.  The excitation laser showed slight mode-hopping; if improved, the resolution should be considerably below 2 GHz. This cavity-enhanced spectroscopy technique adds an important tool to the repertoire for resonant single quantum dot spectroscopy \cite{2008.APL.Warburton.QD-spectroscopy-ac-stark-shift,2003.APL.stark_shift_spectroscopy,2004.PRL.Hogele-Petroff.voltage_controlled_dot,2007.NanoLett.Unlu.QD_resonant_excitation}. Resonance fluorescence from a QD in a cavity was previously reported in a planar optical cavity\cite{2007.PRL.Muller-Shih.QD_resonance_fluorescence_cavity}; however, the excitation geometry used in Ref.\citen{2007.PRL.Muller-Shih.QD_resonance_fluorescence_cavity} is difficult to realize in cavity designs for high Purcell regime or strong coupling, such as photonic crystals or microdisks.  The cavity-enhanced spectroscopy shown here should be applicable for solid state cavity QED systems with many cavity designs, so long as the QD has a large enough pure dephasing rate to drive the cavity. Fig. \ref{fig:coherent_CW}(e) plots the integrated QD intensity as the laser is scanned over the cavity resonance.  The QD dot then emits with a linewidth that appears limited by the our spectrometer resolution of (0.03 \unit{nm}).

\begin{figure}
\centering{\includegraphics[width=.8\linewidth]{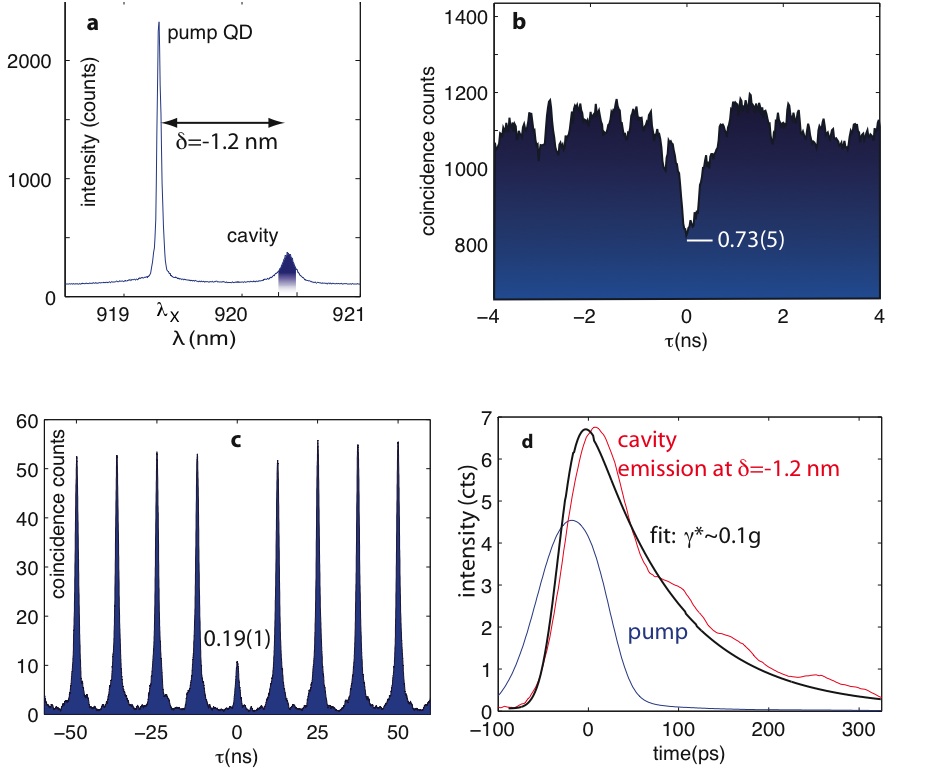}}
\caption{{\footnotesize (a) Resonant QD excitation at 10K. For the autocorrelation measurements, the cavity emission is filtered to a width of 0.2\unit{nm}. (b) Autocorrelation histogram of cavity emission when the QD is pumped resonantly at a power of $12$\unit{nW} in cw mode. (c) Pulsed autocorrelation measurement. (d) Time-resolved resonant pumping of the quantum dot and emission into the cavity mode. A theoretical fit to the cavity emission yields an estimate of the pure dephasing rate.  }}
 \label{fig:g2}
\end{figure}

The mechanism that allows the quantum dot to drive the far off-resonant cavity is not yet completely clear. It has previously been reported that quantum dots that were pumped through higher excited states or the QD wetting layer can drive the cavity even when it is far detuned\cite{2007.Nature.Hennessy-Imamoglu.Strong_coupling_quantum_nature, 2006.PRL.Strauf,2008.PRB.Finley.QD_off_resonant}. Several recent theoretical models attribute the off-resonant driving of the cavity mode to a pure dephasing mechanism of the quantum dot \cite{2008.PRL.Mork.dephasing,2008.OpEx.Noda.dephasing,2008.ArXiv.Poizat.dephasing}. We describe our experimental data with a quantum master equation model that considers the dephasing as an additional Liouvillian $\mathcal{L}_d$ which depends on a dephasing rate $\gamma^{*}$ (see Methods).  The dephasing term allows driving of the cavity (QD) through the QD (cavity), as is shown in the fit in Figs.\ref{fig:coherent_CW}(c,d), where we used $\gamma^{*}=0.1g$. This value of $\gamma^{*}$ was measured independently (see Fig.\ref{fig:g2}(f)) and agrees with values cited in the literature for resonant excitation studies \cite{2004.PRB.Langbein_Wieck.dephasing_400ps}. 

To explore the effect of pumping the off-resonant dot, we measured the cavity emission at various detunings of the QD exciton. At each cryostat temperature given in Fig.\ref{fig:coherent_CW}(h), we tune the cw excitation laser to the QD, keeping the power constant at $50\pm0.5$\unit{nW} before the lens. The integrated cavity intensity is plotted with detuning $\delta=\omega_{cav}-\omega_{qd}$. In our model, we assume a temperature-dependent dephasing\cite{2007.PRB.Favero-Gerard.QD_dephasing_temp} rate $\gamma=\gamma_0+\alpha_0T$, with $\alpha_0=0.5\mu$ eV K$^{-1}$ and $\gamma_0=\kappa/100$. The theory does not fully explain the observation, suggesting that pure dephasing is only a part of the off-resonant driving mechanism between the QD and cavity. Phonon-mediated and two-photon absorption processes probably also play a role, but are not captured in our model.

We now explore the pump power dependence of coherent excitation. Fig.\ref{fig:coherent_CW}(f) shows the spectrometer intensity at $P_{in} = 200 $\unit{nW}. One striking difference is that the features are far more blurred; we believe this results in part because of increased spectral diffusion at high intensity\cite{2008.NPhys.Gerard.motional_narrowing,2007.PRB.Gerard.QD_zero_phonon_linewidth}. A careful scan across the QD, where the tail of the excitation laser is subtracted, gives the cavity emission spectrum shown in Fig.\ref{fig:coherent_CW}(g). The scan reveals a second peak when the pump is tuned to $\lambda_{p}=919.85$\unit{nm}. In Fig.\ref{fig:coherent_CW}(h), we plot the cavity emission as a function of excitation power when $\lambda_p$ is tuned to the lines one by one. The single-exciton line shows a linear pump dependence, as expected.  The second line shows a quadratic pump dependence, which suggests a bi-exciton state that is resonantly pumped by two-photon absorption. The power of this line was too low to confirm this identification by a cross-correlation measurement with the single exciton emission. 

We next turn to measurements of the second-order correlation function $g^{(2)}(\tau)$ to study the quantum nature of the QD/cavity system. We estimate $g^{(2)}(\tau)$ by a measurement of the autocorrelation using a Hanbury-Brown-Twiss (HBT) setup. With the half wave plate in Fig.\ref{fig:setup}(a) aligned to maximize the cavity emission, we obtain a high isolation of the pump, as shown in Fig.\ref{fig:g2}(a). Under this polarization setting, we observed only the single-excition absorption line through the cavity emission. We spectrally filter the cavity emission using a $0.2$\unit{nm} grating filter before the emission is sent to the HBT setup.  The autocorrelation histogram of the cavity emission when the QD is resonantly excited is shown in Fig.\ref{fig:g2}(b). The antibunching depth is limited to $g^{(2)}(0)\sim 0.76$ because the $\sim 300$\unit{ps} detector resolution is longer than the excited state lifetime $\tau_0 \approx 118$\unit{ps}, which is independently measured (see below) and is slower than the 17\unit{ps} shown in Fig.\ref{fig:setup}(g) because the QD is detuned from the cavity. 

The detector resolution is less problematic under pulsed excitation. To limit the pump overlap with the cavity emission, we use 40-ps probe pulses at 80 MHz repetition rate.  The center excitation wavelength is resonant with the QD at $919.5$\unit{nm}.  Fig.\ref{fig:g2}(c) shows the autocorrelation histogram of the cavity emission, which indicates $g^{(2)}(0)=0.19(1)$.  We believe the main contribution to counts near $\tau=0$ is due to the tail of the pulsed excitation laser. Because both the excitation and the emission into the cavity are faster than the detector resolution, it is not possible to distinguish them temporally. We believe the $\tau=0$ peak should be significantly lower with better spectral filtering or higher QD-cavity detuning. 

The autocorrelation measurements demonstrate the use of the resonantly driven QD as an on-demand single photon source. Such a single photon source has several advantages over previously reported QD-based single photon sources. The coherent driving mechanism eliminates timing jitter which results from the random relaxation time of electron-hole pairs under above-resonant excitation. Furthermore, the emission is stabilized by the cavity frequency. The photon indistinguishability can approach unity, albeit at the cost of efficiency\cite{2008.ArXiv.Poizat.dephasing}. In quantum dots that are incoherently pumped through a higher excited state, the combination of dephasing and timing jitter\cite{2004.PRA.Kiraz.SPS-prospects} appears to limit the mean wavefunction overlap to about $\sim 90$\% for the types of QDs employed here \cite{2005.PhysicaE.Vuckovic}. Since the emission occurs through the cavity, it is easier to match the emission from distant QDs for applications such as remote QD entanglement via interference of single photons emitted\cite{2003.PRL.Simon-Irvine.entanglement_indist_photons,2005.PRA.Childress-Lukin.quantum-repeater-SP}.

We performed time resolved measurements of the emission of the resonantly excited QD into the cavity mode. Fig.\ref{fig:g2}(d) shows the 40-ps pump pulse and cavity emission, measured simultaneously on a streak camera. The QD-driven cavity emission lifetime is $\tau \sim 118$\unit{ps} when the dot is detuned by $\delta = -1.2$\unit{nm}. We can use this lifetime measurement to infer the dephasing rate.  We fit the decay time by a Monte Carlo simulation of the master equation, where the QD begins in the ground state and is driven by the 40-ps pump pulse (see Methods). All parameters except for $\gamma^{*}$ are fixed as before.  The fit gives $\gamma^{*}=0.10(1) g$. 

We have studied coherent excitation of a strongly coupled QD/photonic crystal cavity system, with three main results. First, time-resolved reflectivity measurements on the QD/cavity show the vacuum Rabi oscillation of the dot in the cavity and enable a direct means for observing and manipulating the QD. Second, we considered the resonant driving of a cavity-detuned dot which efficiently populates the cavity mode, adding insight to previous non-resonant studies of this phenomenon\cite{2007.Nature.Hennessy-Imamoglu.Strong_coupling_quantum_nature, 2006.PRL.Strauf,2008.PRB.Finley.QD_off_resonant}. This cavity-controlled read-out channel allows high-resolution, resonant single quantum dot spectroscopy. Third, we demonstrated an on-demand single photon source relying on a resonantly driven quantum dot. This source promises unity indistinguishability\cite{2008.ArXiv.Poizat.dephasing}.  Since the emission frequency is set by the cavity resonance, which is easier to control than the inhomogenously distributed QD frequency, this source is furthermore appealing for creating entanglement by photon interference\cite{2003.PRL.Simon-Irvine.entanglement_indist_photons,2005.PRA.Childress-Lukin.quantum-repeater-SP}.   In the future, much larger coupling efficiencies will be required. The all-optical techniques discussed here are compatible with integrated photonic crystal structures, where cavities coupled to single quantum dots may be connected through networks of waveguides\cite{2007.OpEx.Englund,2008.OpEx.Faraon.DIT-WG,2006.OpEx.Noda.efficient_drop_filter,2006.NPhoton.Notomi.trapping_delaying_high_Q} and other chip-integrated elements.


\begin{addendum}
 \item We thank Hideo Mabuchi for helpful discussions. Financial support was provided by the Office of Naval Research (PECASE and ONR Young Investigator
awards), National Science Foundation, Army Research Office, and DARPA Young Faculty Award. A.M. was supported by the SGF (Texas Instruments Fellow). Work was performed in part at the Stanford Nanofabrication Facility of NNIN supported by the National Science Foundation. 
 
 \item[Competing Interests] The authors declare that they have no competing financial interests.
\item[Correspondence] Correspondence and requests for materials should be addressed to Dirk Englund ~(email: englund@stanford.edu).
\end{addendum}

\section*{Methodology}
\subsection{HBT} In autocorrelation measurements, light is filtered to 0.2\unit{nm}, directed through a 50:50 beam splitter, and coupled through multi mode fibers to single photon counter modules. Coincidences are recorded on a time interval analyzer. 

\subsection{Time-domain dynamics of QD/cavity system} 

Fig.\ref{fig:time_domain}: We reflect $40$\unit{ps} pulses from the cavity. The temperature of the cryostat is adjusted to tune the single exciton transition into the cavity.  The spectral and spatial alignment are optimized with the scanning cw laser in the cross-polarized arrangement shown in Fig.\ref{fig:setup}(a). The reflected beam is now recorded on a Hamatsu C5680 streak camera.  

A quantum Monte Carlo simulation based on Ref.\citen{TanMATLAB} is used to model the time-dependent reflectivity.  The values for $g,\kappa,\gamma,\Delta\lambda$ are taken from the spectral characterization. A Gaussian classical field $E(t)=E_0 \exp(-t^{2}/2\sigma_t^{2})$ drives the cavity mode with a FWHM of 40\unit{ps}. The QD is initialized into the ground state $\ket{g}$. With all other system parameters fixed, the amplitude $E_0$ is adjusted for the experiment with lowest power of $P_{in}=0.1$nW. This calibration fixes $E_0$ for experiments with higher power.  Each simulation yields the time-dependent cavity photon number $\braket{a(t^{\dagger}(t)a(t)}$.

From the reflectivity in Fig.\ref{fig:setup}(f), we estimate the QD has a probability of being in an optically dark state of $p_D\approx 0.2$, meaning that a background signal corresponding to an empty cavity reflectivity must be added, giving $\braket{a^{\dagger}(t)a(t)}'=p_{D} \braket{a^{\dagger}(t)a(t)}_{g=0} + (1-p_{D}) \braket{a^{\dagger}(t)a(t)}$, where we set $g=0$ for the first term and $p_D=0.2$. The final fit is obtained by convolving $\braket{a^{\dagger}(t)a(t)}'$ with the streak camera response of $3$\unit{ps}. 

Fig.\ref{fig:setup}(g): To model the PL when the QD is tuned to the cavity and excited incoherently through a higher energy level, we use the Monte Carlo simulation described above, but initialize the QD in the excited state $\ket{e}$. The time jitter is modeled by finding a weighted average of relaxation times, with a 1/e time of $10$ps. 

Fig.\ref{fig:g2}(d): In the Monte Carlo simulation, the QD is detuned by $\delta$ and driven resonantly. The QD is initialized into the ground state and the single-excition transition is driven by the classical field.  The Hamiltonian now has a driving term $E(t) (\sigma_{+}+\sigma_{-})$, where $\sigma_{+,-}$ are the raising and lowering operators of the quantum dot.

\section*{Supplemental Material}

\subsection{Dephasing model}
The Master equation describing a QD (lowering operator
$\sigma=\ket g \bra e$) coupled to a cavity mode (with
annihilation operator $a$) is given by
\begin{equation}
\label{Maseq} \frac{d\rho}{dt}=-i[H,\rho]+\frac{\kappa}{2}(2a\rho
a^\dag-a^\dag a \rho-\rho a^\dag a)+\frac{\gamma}{2}(2\sigma\rho
\sigma^\dag-\sigma^\dag \sigma \rho-\rho \sigma^\dag
\sigma)+\frac{\gamma_{d}}{2}(\sigma_z\rho\sigma_z-\rho)
\end{equation}
where $\gamma$, $\kappa$ and $\gamma_d$ accounts for QD population
decay, cavity population decay and QD pure dephasing;
$\sigma_z=[\sigma^\dag,\sigma]$. $H$ is the hamiltonian of the
system without considering the losses and is given by
\begin{equation}
H=\frac{\Delta}{2}a^\dag a -\frac{\Delta}{2}(\sigma_z)+ig(\sigma
a^\dag-a\sigma^\dag)
\end{equation}
$\Delta$ is QD-cavity detuning and the reference energy is the
mean of QD and cavity energy. Using eq. \ref{Maseq} one can write
\cite{2008.ArXiv.Poizat.dephasing}
\begin{eqnarray}
  \left \langle \frac{da}{dt} \right \rangle = (-i\frac{\Delta}{2}-\frac{\kappa}{2})\langle a \rangle+g\langle \sigma \rangle\\
  \left \langle \frac{d\sigma}{dt} \right \rangle = (i\frac{\Delta}{2}-\frac{\gamma}{2}-\gamma_d)\langle \sigma \rangle+g\langle \sigma_z a \rangle\\
\end{eqnarray}
These two coupled equations are solved. Initial condition $a(0)=1;
\sigma(0)=0$ and $a(0)=0;\sigma(0)=1$ correspond to pumping the
cavity and the dot respectively. Let us define
$A=\left(-i\frac{\Delta}{2}-\frac{\kappa}{2}\right)$ and
$B=\left(i\frac{\Delta}{2}-\frac{\gamma}{2}-\gamma_d\right)$ and
$\lambda_-=A+B-\sqrt{(A-B)^2-4g^2}$ ,
$\lambda_+=A+B+\sqrt{(A-B)^2-4g^2}$. When the dot is pumped, the
solution in time domain is given by
\begin{equation}
a(t)=\frac{(e^{\lambda_+t}-e^{\lambda_-t})g}{\sqrt{(A-B)^2-4g^2}}
\end{equation}
and
\begin{equation}
\sigma(t)=\frac{(A-B)(e^{\lambda_-t}-e^{\lambda_+t})+\sqrt{(A-B)^2-4g^2}(e^{\lambda_-t}+e^{\lambda_+t})}{2\sqrt{(A-B)^2-4g^2}}
\end{equation}
So the spectrum emitted by the cavity $(S_{cav})$ and by the QD
$(S_{QD})$ are given by (by virtue of Quantum Regression theorem):
\begin{equation}
S_{cav}(\omega)=\frac{g^2}{(A-B)^2-4g^2}\left|\frac{1}{\omega-\lambda_-}-\frac{1}{\omega-\lambda_+}\right|^2
\end{equation}
\begin{equation}
S_{QD}(\omega)=\frac{1}{2((A-B)^2-4g^2)}\left|\frac{A-B+\sqrt{(A-B)^2-4g^2}}{\omega-\lambda_-}-\frac{A-B-\sqrt{(A-B)^2-4g^2}}{\omega-\lambda_+}\right|^2
\end{equation}
Similarly, if the cavity is pumped, then depending on the other
initial condition one can solve and get:
\begin{equation}
a(t)=\frac{(B-A)(e^{\lambda_-t}-e^{\lambda_+t})+\sqrt{(A-B)^2-4g^2}(e^{\lambda_-t}+e^{\lambda_+t})}{2\sqrt{(A-B)^2-4g^2}}
\end{equation}
\begin{equation}
\sigma(t)=\frac{(e^{\lambda_-t}-e^{\lambda_+t})g}{\sqrt{(A-B)^2-4g^2}}
\end{equation}
So the spectrum becomes:
\begin{equation}
S_{cav}(\omega)=\frac{1}{2((A-B)^2-4g^2)}\left|\frac{B-A+\sqrt{(A-B)^2-4g^2}}{\omega-\lambda_-}-\frac{B-A-\sqrt{(A-B)^2-4g^2}}{\omega-\lambda_+}\right|^2
\end{equation}
\begin{equation}
S_{QD}(\omega)=\frac{g^2}{(A-B)^2-4g^2}\left|\frac{1}{\omega-\lambda_-}-\frac{1}{\omega-\lambda_+}\right|^2
\end{equation}
In our experimental setup as the collection efficiency from cavity
is much higher compared to QD, hence we can assume that in
experiment what we are observing is $S_{cav}(\omega)$.



\clearpage

\end{document}